\documentclass[twocolumn]{emulateapj}

\usepackage{graphicx}
 \usepackage{amsmath}
\usepackage{txfonts}
\usepackage{longtable}
\usepackage{lscape}
\usepackage{natbib}
\usepackage{soul}

\makeatletter
\AtBeginDocument{\let\LS@rot\@undefined}
\makeatother

\usepackage[usenames,dvipsnames,svgnames,table]{xcolor}

\usepackage{mathtools}
\DeclarePairedDelimiter\bra{\langle}{\rvert}
\DeclarePairedDelimiter\ket{\lvert}{\rangle}
\DeclarePairedDelimiterX\braket[2]{\langle}{\rangle}{#1 \delimsize\vert #2}

\shorttitle{Magnetic sensitivity in the wing scattering polarization signals of H~{\sc i} Ly$\alpha$}%

\shortauthors{Alsina Ballester, Belluzzi \& Trujillo Bueno}

\begin{document}

\title{Magnetic sensitivity in the wing scattering polarization signals \\ of the hydrogen Lyman-$\alpha$ line of the solar disk radiation}

\author{E. Alsina Ballester\altaffilmark{1}, 
L. Belluzzi\altaffilmark{1,2}, and 
J. Trujillo Bueno\altaffilmark{3,4,5}} 

\altaffiltext{1}{Istituto Ricerche Solari Locarno, CH-6605 Locarno Monti, 
Switzerland}
\altaffiltext{2}{Kiepenheuer-Institut f\"ur Sonnenphysik, D-79104 Freiburg, 
Germany}
\altaffiltext{3}{Instituto de Astrof\'{\i}sica de Canarias, E-38205  
La Laguna, Tenerife, Spain}
\altaffiltext{4}{Departamento de Astrof\'{\i}sica, 
Universidad de La Laguna, E-38206 La Laguna, Tenerife, Spain}
\altaffiltext{5}{Consejo Superior de Investigaciones Cient\'{\i}ficas, Spain}

\email{ernest@irsol.ch}

\begin{abstract}
The linear polarization produced by scattering processes in the hydrogen Ly$\alpha$ line of the solar disk radiation
is a key observable for probing the chromosphere-corona transition region (TR) and the underlying chromospheric plasma.
While the line-center signal encodes information on the magnetic field and {the} 
 three-dimensional structure of the TR, 
the sizable scattering polarization signals that the joint action of partial frequency redistribution and $J$-state 
interference produce 
in the Ly$\alpha$ wings have generally been thought to be sensitive only to the thermal structure of the solar 
atmosphere. 
Here we show that the wings of the $Q/I$ and $U/I$ scattering polarization profiles of 
this line are actually sensitive to the presence of chromospheric magnetic fields, with strengths similar to those 
that produce the Hanle effect in the line core {(i.e., {between 5 and} 100 gauss, approximately).}
In spite of the fact that the Zeeman splitting induced by such weak 
fields is very small compared to the total width of the line, 
the magneto-optical effects that couple the transfer equations for Stokes $Q$ and $U$ are actually able to produce 
sizable changes in the $Q/I$ and $U/I$ wings. 
{We find that magnetic fields with longitudinal components {larger than} $100$~G produce an almost complete depolarization of the wings of 
the Ly$\alpha$ $Q/I$ profiles within a ${\pm}5$~\AA\ spectral range around line center, while stronger fields are required for the $U/I$ wing signals to be depolarized to a similar extent. }
The theoretical results presented here further expand the diagnostic content of the 
unprecedented spectropolarimetric observations provided by the Chromospheric Lyman-Alpha Spectropolarimeter (CLASP). 
\end{abstract}


\section{Introduction}
\label{Sect:Introd}

The linear polarization produced by scattering processes in ultraviolet (UV) resonance lines 
of the solar disk radiation encodes key information on the plasma of the upper solar chromosphere and transition region (TR). 
For example, it is known that the line-center scattering polarization signals are sensitive to magnetic fields via the Hanle effect 
\citep[e.g.,][hereafter LL04]{LandiLandolfi04}. 
Of particular interest is the hydrogen Ly$\alpha$ resonance line at $121.6$~nm, 
the strongest emission line in the solar UV spectrum. A few years ago, the Chromospheric Lyman-Alpha 
Spectro-Polarimeter (CLASP) sounding rocket experiment, motivated by theoretical predictions based on radiative transfer (RT) calculations
 \citep{TrujilloBueno+11,Belluzzi+12,Stepan+15}, discovered conspicuous scattering polarization signals in Ly$\alpha$ \citep[see][]{Kano+17}. 
Theoretical modeling of the observed Stokes $Q/I$ and $U/I$ line-center signals recently allowed us to constrain   
the magnetic field strength and geometrical complexity of the corrugated surface that delineates the 
chromosphere-corona TR 
\citep{TrujilloBueno+18}.

While the line-center photons of the hydrogen Ly$\alpha$ line stem mainly from the TR, 
the wing photons encode information on the underlying chromospheric layers (e.g., at $\Delta{\lambda}={\pm}1$ \AA\ from the 
line center, the height in the solar atmosphere where the optical depth is unity lies a few hundred kilometers below the TR).
Unlike the $Q/I$ and $U/I$ line-center signals, which are sensitive to the presence of magnetic fields in the TR via the Hanle effect,  
the wing signals have always been thought to be sensitive only to the thermal structure of the solar 
atmosphere 
\citep[e.g.,][]{Belluzzi+12}. The main aim of the present {paper} is to show that the wings of the $Q/I$ and $U/I$ profiles of the 
hydrogen Ly$\alpha$ line are sensitive to the presence of magnetic fields in the solar chromosphere, with strengths similar to 
those that {characterize} the onset of the Hanle effect in the line core. 
The physical mechanism at the origin of this magnetic sensitivity is as follows. 

In some resonance lines for which the effects of partial frequency redistribution (PRD) produce
large $Q/I$ wing signals, the $\rho_V\,U$ and $\rho_V\,Q$ magneto-optical (MO) terms of the transfer equations for Stokes $Q$ and $U$, 
respectively, can {induce} a significant magnetic sensitivity in the line's scattering polarization 
wings. 
Given that in the line wings $\rho_V$ is significant already for relatively 
weak magnetic fields, the above-mentioned $\rho_V\,Q$ term introduces sizable, magnetically sensitive, $U/I$ wing signals. 
In turn, such large $U/I$ wing signals allow the $\rho_V\,U$ term to introduce a magnetic sensitivity 
in the $Q/I$ wing signals. 
{This mechanism causes both a rotation of the plane of linear polarization as the radiation travels 
through the solar atmosphere \citep[see][]{AlsinaBallester+17}   
and an effective decrease of the degree of total linear polarization \citep[see][]{AlsinaBallester+18}.} 
{Recent} RT investigations have indicated that such MO effects should play an important role in the wings of many
strong chromospheric lines, such as the Mg~{\sc ii} k line \citep{AlsinaBallester+16}, the 
Mg~{\sc ii} h \& k lines \citep{delPinoAleman+16}, the Sr~{\sc ii} $407.8$~nm line \citep{AlsinaBallester+17}, 
 and the Ca~{\sc i} $422.7$~nm line \citep{AlsinaBallester+18}.

Although the physical mechanism that introduces magnetic sensitivity in the Ly$\alpha$ scattering polarization wings 
is therefore not new, it is remarkable that it is capable of producing measurable effects even in a far UV line like 
hydrogen Ly$\alpha$. 
This is because, {at line-wing wavelengths}, the $\rho_V$ coefficient takes sizeable values relative to the absorption coefficient 
already when the Zeeman splitting becomes comparable to the radiative and collisional line broadening. In contrast, the signals produced 
by the familiar Zeeman effect depend on the ratio of the magnetic splitting over the Doppler width of the line and 
therefore scale with the wavelength of the spectral line under consideration. 

\section{Formulation of the problem}
We present the results of non-local thermodynamic equilibrium (NLTE) RT 
calculations of the intensity and linear polarization of the hydrogen Ly$\alpha$ line, 
considering the semi-empirical model C of \cite{Fontenla+93}, hereafter FAL-C. 
The use of this static one-dimensional (1D) solar atmospheric model 
allows us to isolate the 
influence of the magnetic field (although neglecting its possible horizontal fluctuations) from other possible symmetry-breaking mechanisms.
The magnetic fields we have imposed in this model atmosphere are deterministic. 
{Hereafter, we specify their direction by their inclination and azimuth, defined as illustrated in}
Figure~1 of \cite{AlsinaBallester+18}. 
The lines of sight (LOSs) for the considered Stokes profiles are specified by $\mu = \cos\theta$, where $\theta$ is the 
heliocentric angle. The positive direction for Stokes $Q$ has been taken along the $Y$ axis (i.e., parallel to the limb 
for all LOSs with $\mu<1$). 
In the calculations presented below, the line-broadening effect of both elastic and inelastic collisions {is} 
taken into account according to the rates presented in LL04 and \cite{PrzybillaButler04}, respectively. The depolarizing effect of the 
former has not been taken into account, after having verified numerically that its impact is negligible for this very strong 
chromospheric line.
 
{The Ly$\alpha$ line is produced by the transition between the hydrogen levels 
$n=1$ and $n=2$. 
Taking the fine structure (FS) of hydrogen into account, and neglecting the
contribution from forbidden transitions (under the electric dipole
approximation), this line {receives contributions} from two FS 
transitions, namely those between the $^2$P$_{1/2}$ and $^2$P$_{3/2}$ FS levels of the $^2$P upper 
term and the $^2$S$_{1/2}$ FS level of the $^2$S lower term (i.e., the ground state).}
It has been established from previous theoretical investigations in the 
unmagnetized case (Belluzzi et al. 2012) that reliable calculations of the 
wing linear polarization of the hydrogen Ly$\alpha$ line must account for 
quantum interference between the $^2$P$_{1/2}$ and $^2$P$_{3/2}$ {upper levels}  
(i.e., $J$-state interference), in addition to PRD effects.
{An atomic model accounting for the various FS transitions between two terms, 
as well as for the quantum interference between different FS $J$-levels belonging to 
the same term, is generally referred to as a two-term atom (see LL04).
A} correct modeling of the wing scattering polarization of the Ly$\alpha$ line 
thus requires considering at least a two-term ($^2$S -- $^2$P) model atom.

\begin{figure*}[]
\centering
\includegraphics[width = \textwidth]{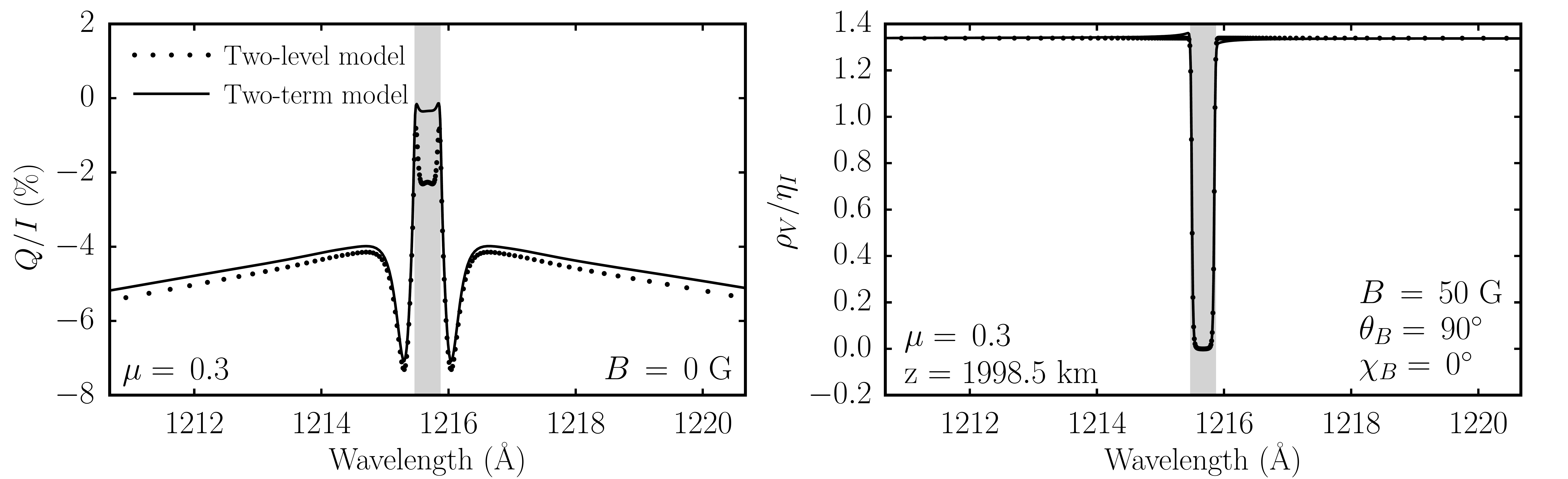}
	\caption{Left panel: $Q/I$ scattering polarization 
	pattern of the hydrogen Ly$\alpha$ line, modeled both as a 
	 two-term atom (solid curve) and as a spinless two-level atom 
	 (dotted curve). 
	Results are obtained from a radiative transfer (RT) calculation, using the FAL-C model, 
	in the absence of a magnetic field.
	Right panel: $\rho_V/\eta_I$ ratio obtained from the same 
	atomic models, in the presence of a $50$~G horizontal magnetic field 
	with azimuth $\chi_B = 0^\circ$, considering the FAL-C model at $1998.5$~km.  
	The gray area indicates the spectral region where 
	the two-level approximation is not suitable.}\label{Fig1} 
\end{figure*}

On the other hand, observing that the FS components are very close to each other, 
it can be shown that, far from the line center, this line behaves in resonance scattering as 
a spinless two-level $0$ -- $1$ transition, in compliance with the 
principle of spectroscopic stability (PSS)\footnote{{The principle of spectroscopic stability is
often stated as follows (see Sect~10.17 of LL04): ``If two different descriptions are used to
characterize a quantum system -- a detailed description which 
takes an inner quantum number into account and a simplified description which disregards it -- the predicted results 
must be the same in all physical experiments where the structure described by the inner quantum 
number is unimportant.''}}. 
{The good agreement between the modeling that accounts for FS and the one that neglects it can} be clearly seen 
in the left panel of Figure~\ref{Fig1}, in which the scattering polarization profiles 
obtained by considering {both a two-term (${}^2$S -- $^2${P}) and a two-level ($0$ -- $1$) model are 
compared}, in the absence of magnetic fields. 
Unless otherwise noted, an LOS with $\mu = 0.3$ is considered in the figures presented {in this work}. 
The expected discrepancy in the line core is a clear 
manifestation of the depolarizing effect of the FS (e.g., LL04). Indeed, 
the gray area across the line-core region, 
{appearing in several of the figures presented in this paper,} 
indicates the spectral interval where the approximation of neglecting FS is not justified. 
The very small deviations found outside the line-core region are due to the approximate treatment
of elastic collisions in the two-term atom calculations \citep[see][]{BelluzziTrujilloBueno14}.

At spectral distances from the line center that are 
much greater than both the Doppler width of the line and 
the magnetic splitting of the energy levels,  
the line emissivity is 
insensitive to both the Hanle and Zeeman effects, provided that the 
collisional broadening is significantly smaller than the natural width of the line \citep[see LL04; also Appendix B of][]{AlsinaBallester+18}. 
{For illustrative purposes, throughout this work we will focus on} {a} wing wavelength {around} which the linear
polarization maximizes.\footnote{ 
The exact spectral position of the maximum of the linear polarization fraction has a slight dependence 
on the LOS and on the magnetic field under consideration.} {More precisely, we consider the 
wavelength at $360$~m\AA\ to the blue of the
line center (hereafter $\lambda_m$) and we point out that} 
this spectral separation is much greater than the magnetic splitting of the energy levels, even in the 
presence of magnetic fields of a few kilogauss. 
%{It is also considerably larger than the Doppler width, which takes a value of approximately $55$~m\AA\ 
%at the atmospheric heights where most of the radiation at wavelength $\lambda_m$ originates from.}
It is also considerably larger than the Doppler width {corresponding to} the atmospheric {regions} 
where most of the radiation at wavelength $\lambda_m$ {comes} from. {Indeed, considering the FAL-C model, 
the Doppler width is approximately $55$~m\AA\ at $z_m = 1998.5$~km; at this height the optical depth at wavelength 
$\lambda_m$ is close to unity for an LOS with $\mu = 0.3$. }

The magnetic sensitivity of the scattering polarization in the wings of this line 
is instead governed 
by the MO effects quantified by the RT coefficient $\rho_V$. {It is important to note  
that the} impact {of such effects} is only appreciable if another physical mechanism, such 
as scattering processes subject to PRD {phenomena}, produces sizable linear polarization 
signals outside the Dopper core \citep[see][]{AlsinaBallester+17}. 
 Using the two-level atomic model, we have verified that, when artificially setting $\rho_V$ to zero, magnetic fields
with strengths up to $5$~kG have no impact on the line's wing linear polarization. 
In the right panel of Figure~\ref{Fig1}, we compare the ratio of $\rho_V$ 
over the absorption coefficient $\eta_I$ 
obtained from the two-term atom equations 
 to that found for a $0$ -- $1$ two-level atom, in the presence of a horizontal magnetic 
field of 50~G\footnote{The Hanle critical field of the hydrogen Ly$\alpha$ line, i.e., the 
magnetic field strength at which the Zeeman splitting of the level with $J = 3/2$ is equal its 
natural width, is {approximately} $53$~G.}.
% Such calculations have been carried out at $z = 1998.5$~km in the 
%FAL-C model. This corresponds to a height at which the optical depth is close to unity for an
%LOS with $\mu = 0.3$ at wavelength $\lambda_m$. 
The {results of the} two {calculations, carried out at height $z_m$ in the FAL-C model, present} an excellent agreement,  
confirming the suitability of neglecting FS when modeling the magnetic sensitivity of this line's wing scattering
polarization signals. 
{As shown in Appendix~\ref{App2L}, the far-wing value of $\rho_V$ is proportional
to the spectral distance between the centers of gravity of the $\sigma_b$ and $\sigma_r$ components of the line. Thus, 
the above-mentioned agreement is ultimately related to the fact that - in accordance with the PSS - the 
frequency shifts of the centers of gravity of the $\sigma_b$, $\pi$, and $\sigma_r$ components for a two-term atom, 
obtained accounting for the incomplete Paschen-Back (IPB) effect,  
coincide with those of a normal Zeeman triplet, i.e., of a spinless
two-level atomic transition (e.g., Section~3.4 of LL04).} 

\section{The impact of magneto-optical effects}
\label{Sect:MO}
{In this} section, we present the results of illustrative 
RT calculations of the Ly$\alpha$ wing scattering polarization signals, 
considering a spinless two-level model atom and 
accounting for the joint impact of  
PRD and of magnetic fields through the Hanle, Zeeman, and MO effects.
Details on the theoretical and numerical  
framework can be found in \cite{AlsinaBallester+17}. 
{We} consider magnetic fields with a constant strength and orientation throughout  
 the FAL-C model atmosphere, paying particular attention to 
{vertical and horizontal (as well as nearly vertical and {nearly} horizontal)} magnetic fields. 
 
\subsection{Linear Polarization Profiles for Deterministic Magnetic Fields} 
 \begin{figure*}[!t]
 \centering
 \includegraphics[width=\textwidth]{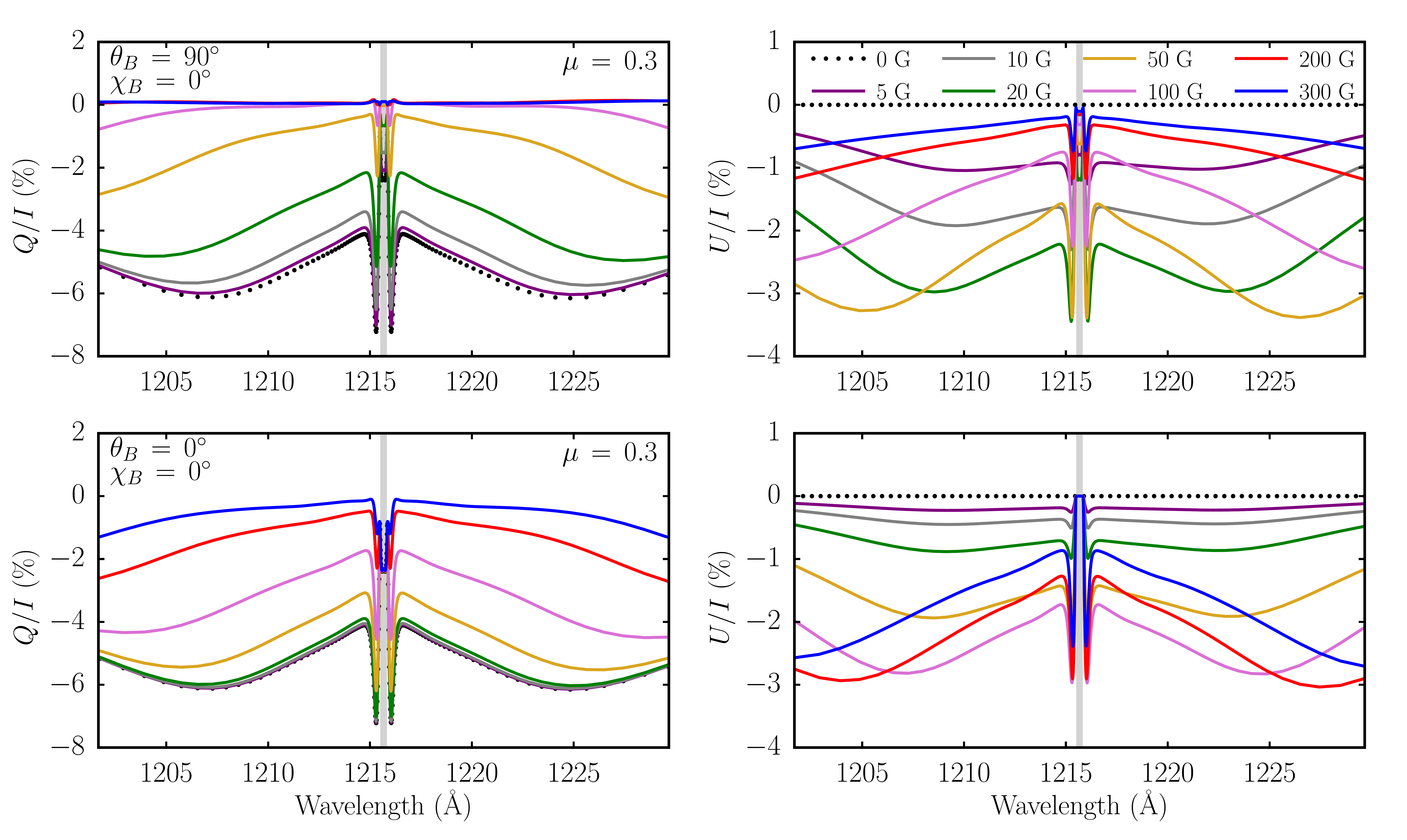}
\caption{The Stokes $Q/I$ {(left panels)} and $U/I$ {(right panels)}, calculated 
 {in} the presence of both horizontal ($\theta_B = 90^\circ$) magnetic fields with azimuth
 $\chi_B = 0^\circ$ {(top panels)} and vertical ($\theta_B = 0^\circ$) magnetic fields 
 {(bottom panels)}.  The colored curves ({see the legend}) correspond to the various considered field strengths. 
 The black dotted curves represent the unmagnetized reference case.
 {When considering horizontal magnetic fields with the same strengths but with azimuth $\chi_B = 180^\circ$ 
 (not shown), the resulting $Q/I$ profiles coincide exactly with those shown for horizontal magnetic fields with
 $\chi_B = 0^\circ$, while the corresponding curves for $U/I$ are identical in
 absolute value but opposite in sign. Likewise, the $Q/I$ profiles obtained in the presence of vertical downward-pointing
 magnetic fields ($\theta_B = 180^\circ$; not shown) coincide with those wtih $\theta_B = 0^\circ$ and a sign reversal is
 found in the 
 $U/I$ profiles}. }
\label{Fig2} 
\end{figure*}
{The top panels of} Figure~\ref{Fig2} {show} the linear polarization profiles {at an LOS with 
$\mu = 0.3$, in the presence of} {horizontal ($\theta_B = 90^\circ$) magnetic} fields {of} various strengths with {azimuth 
$\chi_B = 0^\circ$} 
{(this choice of azimuth maximizes the longitudinal component of the magnetic field).} 
Outside the Doppler core, the MO effects induced by such magnetic fields produce a $U/I$ signal and a
depolarization in $Q/I$. 
{The influence of such effects is controlled by the ratio of $\rho_V$ over $\eta_I$, which 
depends on the longitudinal component of the magnetic field.
Interestingly, this ratio scales with the same parameters
that characterize the efficacy of the Hanle effect (see Appendix~\ref{App2L}). Indeed, such MO effects 
are expected to noticeably impact the wings of the linear polarization signals when the magnetic field strength is 
comparable to the Hanle critical field. 
Moreover, in the presence of increasingly strong magnetic fields, the impact of such MO effects 
is appreciable in the wings of both $Q/I$ and $U/I$ at greater spectral distances from the line 
center. As pointed out in \cite{AlsinaBallester+18}, 
the relative contribution of continuum processes to $\eta_I$ is greater farther into the line wings, 
implying that stronger magnetic fields are required in order for the $\rho_V/\eta_I$ 
ratio to be significant. 
Note also that, in} 
 addition to their amplitude, also the sign of the $U/I$ wing signals is sensitive to the 
{orientation} of the magnetic field. 
{For instance, comparing horizontal magnetic fields with $\chi_B = 0^\circ$ and $\chi_B = 180^\circ$, 
which have longitudinal components of the same magnitude but point in the opposite direction, we have verified that 
the  depolarization of $Q/I$ is the same, 
while the resulting $U/I$ wing signal is identical in absolute value but with opposite sign.} 
{We point out that, because the wing $Q/I$ scattering polarization signals are negative in the unmagnetized case, 
the MO effects induced by a magnetic field with a positive (negative) longitudinal component give rise to 
negative (positive) $U/I$ signals.}
 
{We have also considered the case of vertical magnetic fields ($\theta_B = 0^\circ$) of increasing
strength, for an LOS with $\mu = 0.3$.  
As seen in the bottom left panel of Figure~\ref{Fig2}, for $B = 300$~G 
the wings of $Q/I$ are almost completely depolarized within a ${\pm}5$~\AA\ spectral range around line center, 
 and a significant depolarization is also appreciable much farther into the wings.
These profiles have a strong resemblance to those obtained in the presence of a horizontal magnetic
field of $100$~G discussed above. This can be easily understood by observing that the longitudinal components of
the two aforementioned field configurations are very similar  
(around $90$~G). 
Interestingly, in the presence of magnetic fields with such longitudinal components, 
the near wings of the $U/I$ profiles  
still have a considerable amplitude
(see the right panels of Figure~\ref{Fig2}), 
and stronger magnetic fields are required in order for them to be considerably depolarized. 
Indeed, we have checked that the absolute value of the $U/I$ wing signal at $\lambda_m$ does not fall below $0.1$\%
until magnetic fields with longitudinal components larger than $900$~G are considered. 
Finally, just as in the case of a horizontal magnetic field, we have also verified that if the  
vertical magnetic field is oriented in the opposite direction (i.e., $\theta_B = 180^\circ$), the resulting depolarization of $Q/I$ is the same 
and the $U/I$ profile has the opposite sign, again as a consequence of the sign reversal of the LOS projection of the magnetic field.}

{The previously discussed signatures of the MO effects, namely the depolarization of $Q/I$ together with the 
 appearance of a $U/I$ 
signal whose sign depends on the orientation of the magnetic field, offer} a 
new tool for inferring the longitudinal component of the magnetic fields in the 
chromospheric regions where the Ly$\alpha$ wings originate. 
The magnetic sensitivity of this line's wing scattering polarization signals 
can be expected to be well above the noise level, {even} in quiet regions of the solar atmosphere
{where} the circular polarization signals produced by the Zeeman effect would be extremely weak. 
Moreover, the scattering polarization signal is clearly appreciable 
very far into the line wings, thus encoding information on the magnetic activity in deeper chromospheric layers. 

\subsection{Center-to-limb Variation for Determinstic Magnetic Fields}
\begin{figure*}[!h]
\centering
\includegraphics[width = \textwidth]{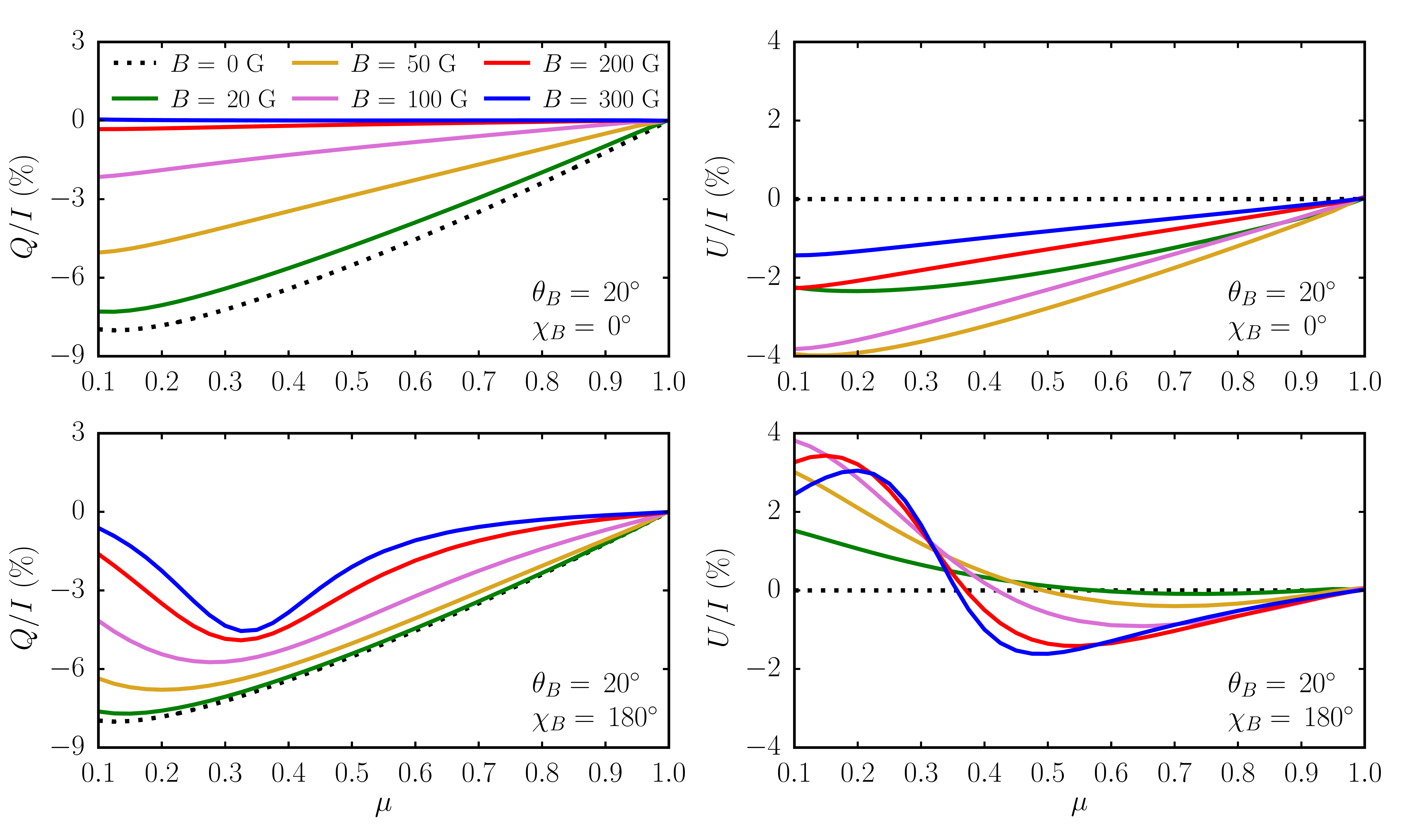}
\caption{{Center-to-limb variation (CLV) of the $Q/I$ (left panels) and $U/I$ (right panels) wing signals 
obtained at $360$~m\AA\ to the blue of line center (i.e., the $\lambda_m$ wavelength defined in the text), 
for magnetic fields with inclination $\theta_B = 20^\circ$ and azimuths $\chi_B = 0^\circ$ (top panels) and 
$\chi_B = 180^\circ$ (bottom panels). The colored curves (see the legend) correspond to various field strengths up to $300$~G. 
The black dotted curves represent the reference unmagnetized case.}}
\label{Fig4a}
\end{figure*}
\begin{figure*}[!h]
\centering
\includegraphics[width = \textwidth]{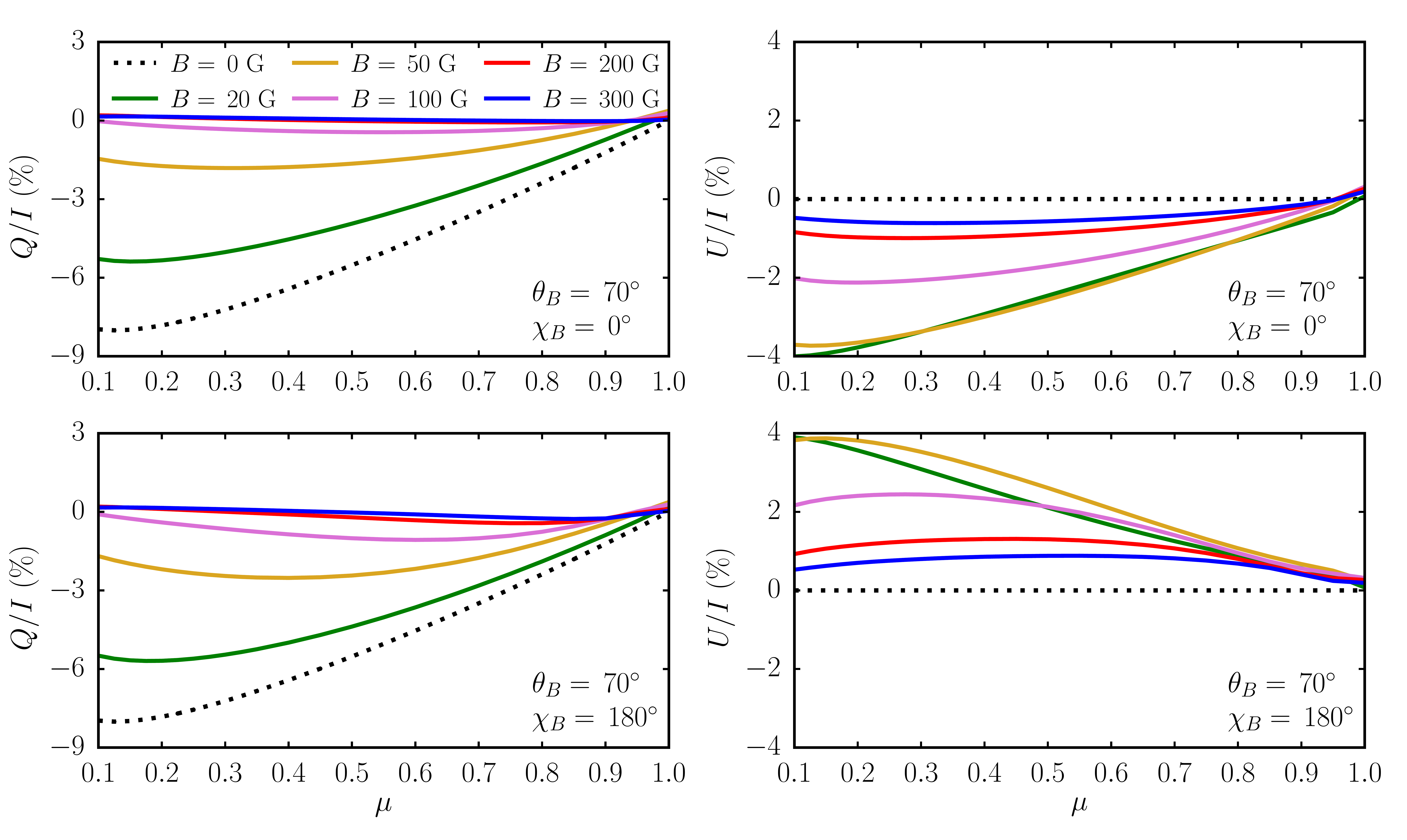}
\caption{{Same as the previous figure, but in the presence of magnetic fields with inclination $\theta_B = 70^\circ$.}}
\label{Fig4b}
\end{figure*}
\begin{figure*}[!t]
\centering
\includegraphics[width = \textwidth]{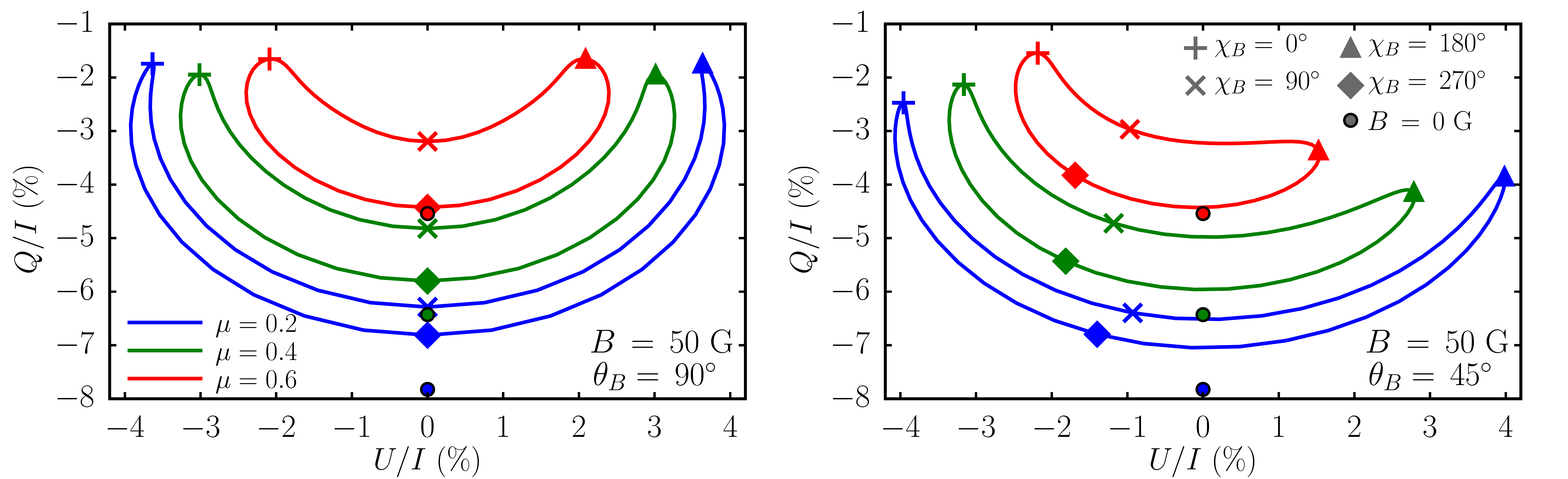}
\caption{Polarization diagrams for 
magnetic fields of $50$~G, with inclinations of $\theta_B = 90^\circ$ (left panel) and $45^\circ$ (right panel). Each closed curve
represents the change in scattering polarization with azimuth, for LOS with $\mu = 0.2$ (blue curves), $\mu = 0.4$ (green curves),
and $\mu = 0.6$ (red curves). The various markers indicate specific azimuths (see legend), except for the circles, which represent the 
unmagnetized reference case.}
\label{Fig5}
\end{figure*} 
{Figures~\ref{Fig4a} and \ref{Fig4b} show} the center-to-limb variation (CLV) of the $Q/I$ and $U/I$ signals at $\lambda_m$, 
%sensitive to MO effects, 
calculated for {nearly vertical ($\theta_B = 20^\circ$; Figure~\ref{Fig4a}) and nearly 
horizontal ($\theta_B = 70^\circ$; Figure~\ref{Fig4b}) magnetic fields of various strengths up to $300$~G, both for  
$\chi_B = 0^\circ$ and $\chi_B = 180^\circ$.

The CLV for $Q/I$ and $U/I$ found in the presence of magnetic fields with $\theta_B = 20^\circ$ and $\chi_B = 0^\circ$
can be explained in a relatively straightforward manner. The projection of the magnetic field along the LOS -- and therefore the
value of $\rho_V/\eta_I$ -- has the same sign for all $\mu$ between $0$ and $1$ and increases monotonically up to $\mu \approx 0.94$.
As the magnetic field strength increases, one finds 
a progressively greater departure} from the $(1 - \mu^2)$ trend for $Q/I$, theoretically predicted in the unmagnetized case. 
The amplitude of the $U/I$ signals, produced by the same MO effects, is found to decrease monotonically with $\mu$, 
 because it depends on both the longitudinal component of the magnetic field and the amplitude of the $Q/I$ signals. 
 {The 
 $U/I$ signals increase in amplitude with the field strength up to roughly $50$~G, but for even stronger fields 
 they  begin to decrease, as the MO effects produce a net reduction of the total fraction of linear polarization 
 \citep[see Appendix A of][]{AlsinaBallester+18}}. 
 
{The situation is substantially different in the presence of magnetic fields with $\theta_B = 20^\circ$ and $\chi_B = 180^\circ$ 
(see the bottom panels of Figure~\ref{Fig4a}). In this case, the magnetic field points away from the observer for LOSs with small $\mu$
values, it becomes completely transversal at $\mu \approx 0.34$, and its longitudinal component becomes 
positive and increases as one continues approaching $\mu = 1$. 
Compared to the case in which $\chi_B = 0^\circ$, the longitudinal component is smaller when considering LOSs with small $\mu$ values, resulting 
in a much more modest depolarization in $Q/I$, especially around the LOS at which the magnetic field is transversal. 
It is interesting to note that, even at this LOS, the magnetic field still produces some depolarization, despite
the fact that $\rho_V$ is zero in this direction. This can be explained because the pumping radiation field is nevertheless 
modified by MO effects, thereby impacting the linear polarization {emitted in} this direction
\citep[e.g.,][]{AlsinaBallester+16,AlsinaBallester+18}. 
There are also clear qualitative differences with respect to the previous case in the CLV for $U/I$; in this case their signals are 
positive for LOS with large inclinations and become negative when directions closer to the vertical are considered. The sign inversion 
occurs around the LOS for which such fields are transversal, although the exact $\mu$ value changes with the field strength because of 
modification of the pumping radiation field induced by MO effects.} 

{On the other hand, when considering nearly horizontal ($\theta_B = 70^\circ$) magnetic fields (see Figure~\ref{Fig4b}), the CLV obtained 
in the presence of fields with $\chi_B = 0^\circ$ and $\chi_B = 180^\circ$ are qualitatively very similar to each other. 
Nevertheless, is worth noting that, for LOSs with small $\mu$ values, the longitudinal component of the magnetic field -- and thus the depolarization of 
$Q/I$ -- is slightly greater for the former case than for the latter. 
For small $\mu$, such nearly horizontal magnetic fields give rise to a considerably stronger depolarization than those
with an inclination of $\theta_B = 20^\circ$, also in this case due to their larger longitudinal components. }

\subsection{A Look at Observational Data: CLASP}
Recently, CLASP successfully measured the linear polarization signals of 
the Ly$\alpha$ line emerging from quiet regions of the Sun, spanning from off-limb positions to close to the disk center 
\citep[see][]{Kano+17}. {In the wings of the $Q/I$ and $U/I$ profiles, considerable fluctuations along the spatial direction 
of the radially oriented slit were found. 
{The} amplitude of the wing $Q/I$ signal was found to decrease 
with $\mu$ (in agreement with our theoretical expectations),
while no {serious} CLV was observed in the amplitude of $U/I$}. 
{We are confident that the observed lack of CLV in the $U/I$ wing signals can be explained by accounting for} 
horizontal variations in the longitudinal component of the magnetic field, 
{which could substantially modify} the net amplitude of the signals resulting from MO effects, {and/or }
by the axial asymmetries in other thermodynamical properties of the solar atmosphere, which may produce $U/I$ signals of
non-magnetic origin. 
An accurate modeling of the {scattering} polarization signals observed {in strong resonance
lines such as H~{\sc i} Ly$\alpha$} must {therefore account} for 
the full three-dimensional complexity of
the solar atmosphere, as well as the joint action of scattering polarization with PRD phenomena and the Hanle, Zeeman, 
and MO effects. 

{In spite of the simplification that the FAL-C semi-empirical model implies, it is 
worthwhile to note that the results of our radiative transfer calculations in this 1D model of the solar atmosphere
can be invoked to qualitatively explain one of the other spectacular observational results provided by CLASP. In addition
to the wavelength variation of the linear polarization profiles, CLASP provided Stokes $I$ and Stokes $Q/I$ broadband images
over a large field of view \citep[see][]{Kano+17}. Within this field of view there was a bright plage and a multitude of network and inter-network
features. Interestingly, the bright plage region and some of the network features that can be distinguished in the Stokes
$I$ image show nearly zero linear polarization in the Stokes $Q/I$ image, while the surrounding quiet regions instead show
very significant polarization signals. In a forthcoming publication we will investigate whether this can be explained on the basis of 
the results reported above, by noting that the broadband $Q/I$ signals observed by CLASP are dominated by the linear polarization
in the Ly$\alpha$ wings \citep[see][]{Belluzzi+12} and by bearing in mind that plages and the network have stronger magnetic fields  
than the surrounding quieter regions.}

\subsection{Unresolved Magnetic Fields} 
Further insights into the magnetic sensitivity of the linear polarization in the wings can be gained by 
studying its behavior on the $Q/I-U/I$ plane.  
The closed curves in the polarization diagrams shown in Figure~\ref{Fig5} indicate how the fractional linear
polarization signals obtained at $\lambda_m$ change 
with $\chi_B$ in the presence of $50$~G magnetic fields with a fixed inclination. 
The diagram is symmetric around the $U/I = 0$ axis for $\theta_B = 90^\circ$ (left panel), but 
this is not the case for arbitrary inclinations, implying the following \citep[see][]{AlsinaBallester+18}.
If one measures the net $U/I$ to be zero in a given spatially unresolved region of the 
solar atmosphere, this is an indication that the magnetic field therein is {transversal, or otherwise} has
a distribution such that the averaged longitudinal component is zero\footnote{ 
From symmetry considerations it can be seen that the two following possible scenarios fulfil
this condition: 
(a) a magnetic field distribution with axial symmetry around a given axis that is perpendicular to the LOS, and 
(b) a distribution with axial symmetry around any given axis, having also reflective symmetry with respect to the plane 
normal to the same axis. The configuration presented in the left panel of Figure~\ref{Fig5} is a particular case of the latter.}. 

{We} also point out that magnetic field {distributions} that do not fulfil the aforementioned condition {are} 
capable of producing a net $U/I$ signal {even if} their orientations change at scales below the {mean free path of the line's photons} 
{(i.e., micro-structured magnetic fields)}. {Indeed, for such a field configuration, in which the inclination is fixed and the 
azimuth changes randomly, the $\rho_V$ is generally not zero \citep[see Equations~(6b) and~(50a) of][]{AlsinaBallester+17}, implying that 
a $U/I$ signal may be produced in the wings. By contrast, for the same field configuration the Hanle effect may modify 
the line-core $Q/I$ amplitude of the scattered radiation, but it produces no $U/I$ signal (see Eqs.~(11), (21), (22), 
and (50b) of the same paper). 
Such qualitative differences can be understood
by realizing that the MO effects quantified by $\rho_V$ depend only on the net longitudinal component of the magnetic field, which
is only zero for all LOSs if $\theta_B = 90^\circ$. On the other hand, the Hanle effect also depends on the angle between the 
magnetic field and the symmetry axis of the pumping radiation field, which in a 1D unmagnetized atmospheric model is parallel 
to the local vertical. The field configuration discussed here is symmetric around this axis and, as a result, the Hanle effect does not 
cause a rotation of the plane of linear polarization of the scattered radiation, although it may decrease the degree of linear 
polarization.}  

\section{Concluding comments}
\label{Sect:Concl} 
 Motivated by the recent theoretical discovery that the wing scattering polarization of some strong resonance lines 
is highly sensitive to the MO effects quantified by the $\rho_V$ terms of the transfer equations for Stokes $Q$ and $U$, we
have conducted an RT investigation on {the} wing linear polarization signals {of} the
hydrogen Ly$\alpha$ line. {We have} modeled {this line considering} a spinless two-level atom {(i.e., the 
impact of FS has been neglected), {having} shown that this approximation is suitable outside the Doppler core.}  
We {have found} that the wing scattering polarization signals of this far UV line are in fact sensitive to longitudinal magnetic fields, 
even {when} they are 
considerably weaker than the Hanle critical field.

Such signals extend far into the line wings, potentially offering a method 
to simultaneously infer the LOS components of the magnetic fields present in {a wide range of depths throughout} the solar chromosphere. 
The sign of such components can be determined from that of $U/I$, while the combined amplitude of $Q/I$ and $U/I$ are indicative of 
their magnitude. From symmetry {considerations applied} to the polarization diagrams, we conclude that measuring a nonzero $U/I$ wing signal may be
a signature of an asymmetry of the distribution of the LOS component of the magnetic field{ within the considered 
spatial resolution element}. 
 
This investigation reveals that relatively weak magnetic fields may strongly impact {the} wing scattering polarization
signals {of the Ly$\alpha$ line} via MO effects. 
{Interestingly, the broadband $Q/I$ images provided by the CLASP suborbital rocket experiment revealed 
{linear polarization} signals {close to zero} in the regions of the field of view corresponding to a plage and to some of the network features, 
in contrast to the much less magnetized surrounding
regions. 
As we shall show in detail in a forthcoming publication, these observations can potentially be explained on the basis of 
the depolarization that MO effects produce in the wings of the Ly$\alpha$ line.} 

{Finally, we emphasize that} an 
accurate RT modeling of the {scattering polarization in the} hydrogen Ly$\alpha$ line requires accounting for the 
3D structure of the solar atmosphere, in addition to the joint action of resonance scattering with PRD and the Hanle,
Zeeman, and MO effects. 

\acknowledgements
E.A.B. and L.B. gratefully acknowledge financial support by the Swiss National Science Foundation (SNSF) through 
Grant 200021\_175997. 
J.T.B. acknowledges the funding received from the European Research Council (ERC) under the European Union's Horizon 2020 
research and innovation programme (ERC Advanced Grant agreement No. 742265).

\appendix
\section{A. The far-wing limit of the elements of the propagation matrix}
\label{AppFarProp}
Here we present an analytical study of
the magnetic dependence of the elements of the line contribution to the so-called propagation matrix (see LL04), 
focusing on spectral regions far beyond the Doppler core. 
We consider a two-term atomic model {without hyperfine structure}, in the presence of magnetic fields of arbitrary strength.
In order to {determine} the various eigenstates of an atomic system in the presence of an external magnetic field, one must
diagonalize the total Hamiltonian $H = H_0 + H_B$, in which $H_0$ is the Hamiltonian of the unperturbed atomic system
and $H_B$ is the magnetic Hamiltonian \citep[see][]{CondonShortley35}. 
Taking the quantization axis of total angular momentum $J$ (i.e., the $z$-axis) parallel to the magnetic field, the 
magnetic Hamiltonian obeys the following commutation rules, 
\begin{equation*}
 [H_B, J_z] = 0 \, , \quad [H_B, J_x] \ne 0 \, , \quad [H_B, J_y] \ne 0 \, .  
\end{equation*}
Therefore, in the presence of a 
magnetic field the quantum number $J$ generally loses {the} property of being a ``good'' quantum number, while this property is 
preserved for the quantum number $M$.
When the magnetic energy is much smaller than the energy intervals of $H_0$ the effect of $H_B$ can be computed
through a perturbative approach to first order \citep[e.g.,][]{LandiDeglInnocenti14}, which implies its diagonalization over the 
degenerate eigenvectors of $H_0$. The matrix $\bra{\beta L S J M}\! |H_B|\!\ket{\beta L S J M^\prime}$ is  
diagonal and the magnetic field produces an energy splitting of the magnetic sublevels that scales linearly with the field strength. 
This approach is commonly known as 
the linear Zeeman splitting approximation (LZS). 
In the more general case, commonly referred to as the IPB effect regime, 
when performing the diagonalization of the total Hamiltonian on the basis 
$\ket{\beta L S J M}$, one finds that the magnetic field produces a 
mixing of the various $J$-levels. The ensuing eigenvectors are  
characterized by quantum number $M$ as well as by the label $j$: 
 \begin{align*}
  & H \, \ket{\beta_u L_u S j_u M_u} = E_{j_u}(\beta_u L_u S, M_u) \, \ket{\beta_u L_u S j_u M_u} \, ; \quad \quad
  \ket{\beta_u L_u S j_u M_u} = \sum_{J_u} C^{j_u}_{J_u}(\beta_u L_u S, M_u) \, \ket{\beta_u L_u S J_u M_u}  \, , \\
  & H \, \ket{\beta_\ell L_\ell S j_\ell M_\ell} = E_{j_\ell}(\beta_\ell L_\ell S, M_\ell) \, \ket{\beta_\ell L_\ell S j_\ell M_\ell} \, ; 
  \quad \quad \; \;   
 \ket{\beta_\ell L_\ell S j_\ell M_\ell} = \sum_{J_\ell} C^{j_\ell}_{J_\ell}(\beta_\ell L_\ell S, M_\ell) \, \ket{\beta_\ell L_\ell S J_\ell M_\ell} \, , 
 \end{align*}
where the $u$ and $\ell$ subscripts refer to the states of the upper and lower term, respectively. $E_j (\beta L S, M)$ is the 
energy for each eigenstate and the $C^j_J(\beta L S, M)$ coefficients describe the coupling between such states
and the $\ket{\beta L S J M}$ basis eigenvectors. Given that the sum of the eigenvalues of a Hamiltonian are equal to its
trace, it can be shown that for each term
\begin{equation*}
 \sum_{j M} E_j(\beta L S, M) = n \, E(\beta L S) \, ,
\end{equation*}
where $n$ is the number of different eigenstates belonging to the considered term and $E(\beta L S)$ is the energy of the term. 
Each of the (electric-dipole) radiative transitions between the various states of the upper term $\ket{\beta_u L_u S j_u M_u}$ and those of the 
lower term $\ket{\beta_\ell L_\ell S j_\ell M_\ell}$ are characterized by their frequencies
\begin{equation}
 \nu_{j_u M_u, j_\ell M_\ell} = \bigl[E_{j_u}(\beta_u L_u S, M_u) - E_{j_\ell}(\beta_\ell L_\ell S, M_\ell)\bigr]/h \, ,
 \label{Freq}
\end{equation}
where $h$ is the Planck constant. %We introduce the reduced frequency shifts 
{These frequencies can also be expressed as shifts with respect to the reference frequency of the multiplet 
 $\nu_0 = \bigl[E(\beta_u L_u S) - E(\beta_\ell L_\ell S)\bigr]/h$, in units of the Doppler width $\Delta\nu_D$, as }  
\begin{equation}
  x_{j_u M_u, j_\ell M_\ell} = \frac{\nu_{j_u M_u, j_\ell M_\ell} - \nu_{0}}{\Delta\nu_D} \, .
  \label{FreqShiftTrans}
\end{equation}
{Introducing also the reduced frequency}
\begin{equation}
x = \frac{\nu_0 - \nu}{\Delta\nu_D} \, , 
\end{equation}
{we note that $x_{j_u M_u, j_\ell M_\ell} + x = (\nu_{j_u M_u, j_\ell M_\ell} - \nu)/\Delta \nu_D$. Moreover, }
it can easily be shown that
 \begin{equation}
  \sum_{j_u M_u j_\ell M_\ell} x_{j_u M_u, j_\ell M_\ell} = 0 \, .  
 \end{equation}
The various transitions between the upper and
lower term can be divided into three groups according to $\Delta M \equiv (M_u - M_\ell) = (\pm 1,0)$. Following the terminology generally used in the literature, we refer to the groups
with $q = - \Delta M = (-1,0,1)$ as the $\sigma_r$, $\pi$, and $\sigma_b$ components, respectively. The strength of each transition is given by 
(see LL04)
\begin{align}
& S^{j_u M_u, j_\ell M_\ell}_q = \frac{3}{2 S + 1} %\notag \\
\sum_{J_u J_u^\prime} C^{j_u}_{J_u}(\beta_u L_u S, M_u) \, C^{j_u}_{J_u^\prime}(\beta_u L_u S, M_u) 
\sum_{J_\ell J_\ell^\prime} C^{j_\ell}_{J_\ell}(\beta_\ell L_\ell S, M_\ell) \, C^{j_\ell}_{J_\ell^\prime}(\beta_\ell L_\ell S, M_\ell) \notag \\
& \times \sqrt{(2 J_u + 1)(2 J_u^\prime + 1)(2 J_\ell + 1)(2 J_\ell^\prime + 1)} %\notag \\
\left\{\begin{array}{c c c}
                J_u   & J_\ell     & 1 \\
                L_\ell    & L_u  & S
                \end{array} \right\}
         \left\{\begin{array}{c c c}
                J_u^\prime      & J_\ell^\prime    & 1 \\
                L_u             & L_\ell     & S
              \end{array} \right\} % \notag \\
         \left(\begin{array}{c c c}
                J_u   & J_\ell  &  1  \\
                -M_u  & M_\ell  & -q
               \end{array} \right)  
          \left(\begin{array}{c c c}
                J_u^\prime   & J_\ell^\prime  &  1  \\
                -M_u  & M_\ell  & -q
               \end{array} \right) \, , 
\label{TransStr}
\end{align}
which fulfil the following normalization condition 
\begin{equation}
 \sum_{j_u M_u j_\ell M_\ell} S^{j_u M_u, j_\ell M_\ell}_q = 1 \, , \; \quad q = (-1,0,1) \, .
\end{equation}

\begin{figure}[!h]
\includegraphics[width=\textwidth]{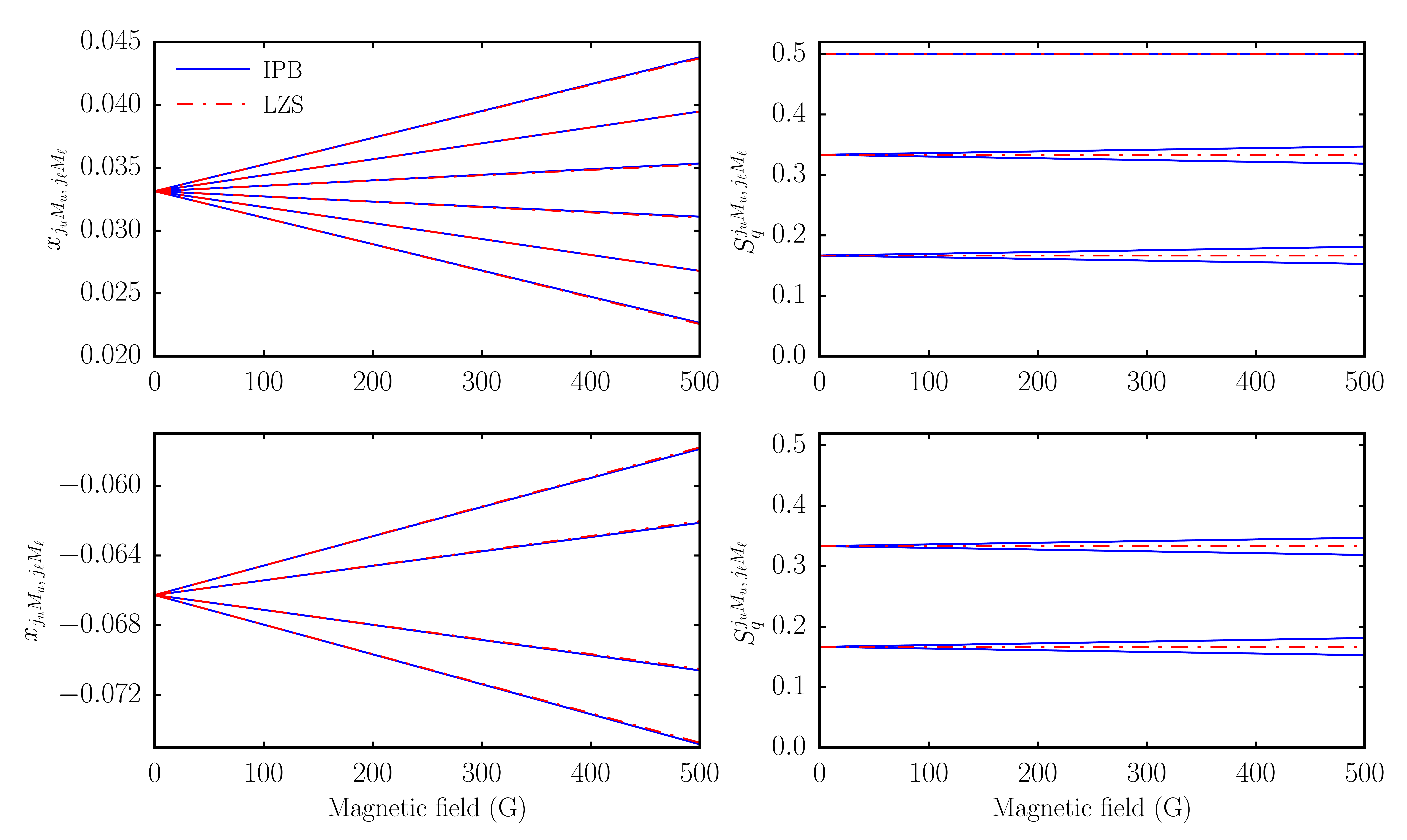}
\caption{Left panels: reduced frequency shifts (see Equation~\eqref{FreqShiftTrans}) for the various transitions between the upper 
 ($^2$P) and lower ($^2$S) term of the Ly$\alpha$ line, as a function of magnetic field strength. 
  The Doppler width has been taken at a height of $1998.5$~km in the FAL-C atmospheric model, corresponding to $54.4$~m\AA . 
Right panels: normalized strengths for the same transitions (see Equation~\eqref{TransStr}), as a function 
  of magnetic field strength. 
The top (bottom) panels illustrate the transitions 
whose upper state has total angular momentum $J_u = 3/2$ ($J_u = 1/2$) in the absence of magnetic field. 
The black solid curves represent the results of the calculation accounting for the incomplete Paschen-Back effect, while 
those represented by the red dashed-dotted curves are obtained under the linear Zeeman splitting approximation. {Note that several of the 
curves corresponding to the strengths of different transitions may overlap.}} 
 \label{FigAppendix1}
\end{figure}
As discussed in this paper, under the electric-dipole approximation the H~{\sc i} Lyman-$\alpha$ line can be modeled
as a two-term atom whose upper term has two FS levels, $^2$P$_{1/2}$ and $^2$P$_{3/2}$. Relative to  
the ground state, their energies are $82258.919$~cm$^{-1}$ and $82259.285$~cm$^{-1}$, respectively. 
Considering field strengths of up to $500$~G, we have verified that the $x_{j_u M_u, j_\ell M_\ell}$ frequency shifts 
calculated making the LZS approximation present a very good agreement with those obtained in the general IPB effect regime, as 
is shown in the left panels of Figure~\ref{FigAppendix1}. 
The quality of this agreement should not be surprising, because the energy separation between the FS levels of the upper term
is more than one order of magnitude larger than the splitting between $M$-levels induced by a magnetic field of such strength. 

On the other hand, we note that when the 
LZS approximation is made, 
the $C^{j}_{J}(\beta L S, M)$ coefficients reduce to $\delta_{J, j}$ and  
the magnetic dependence of the transition strengths given in Equation~\eqref{TransStr} completely vanishes. 
This contrasts with the results obtained in the IPB effect regime, in which the transition strengths are
appreciably modified by such weak magnetic fields, as is shown in the right panels of Figure~\ref{FigAppendix1}. 

The frequencies of the centers of gravity of the $\sigma_b$, $\pi$, and $\sigma_r$ groups, relative to $\nu_0$ 
and in units of Doppler width are defined as 
 \begin{equation}
  \bar{x}_q \, = \sum_{j_u j_\ell M_u M_\ell} S^{j_u M_u, j_\ell M_\ell}_q  x_{j_u M_u, j_\ell M_\ell} \, .
  \label{FrequencyShifts}
 \end{equation}
For the discussions below, it will also be useful to introduce $\bar{\nu}_q \equiv \bar{x}_q \, \Delta\nu_D$. 
It can be shown that, for a two-term atomic model 
in the IPB effect regime, 
the frequency shifts of the centers of gravity of the three groups scale linearly with the strength of 
 the magnetic field (see LL04), according to
\begin{equation}
 \bar{x}_q = -q \, \frac{\nu_L}{\Delta\nu_D} \, ,
 \label{FreqShiftNZT}
\end{equation}
in which we have introduced the Larmor frequency $\nu_L = \mu_0 B/h$, where $\mu_0$ is the Bohr magneton.
Such frequency shifts coincide with those for a spinless two-level atomic model. 
\begin{figure}[!h]
  \centering
 \includegraphics[width=0.68\textwidth]{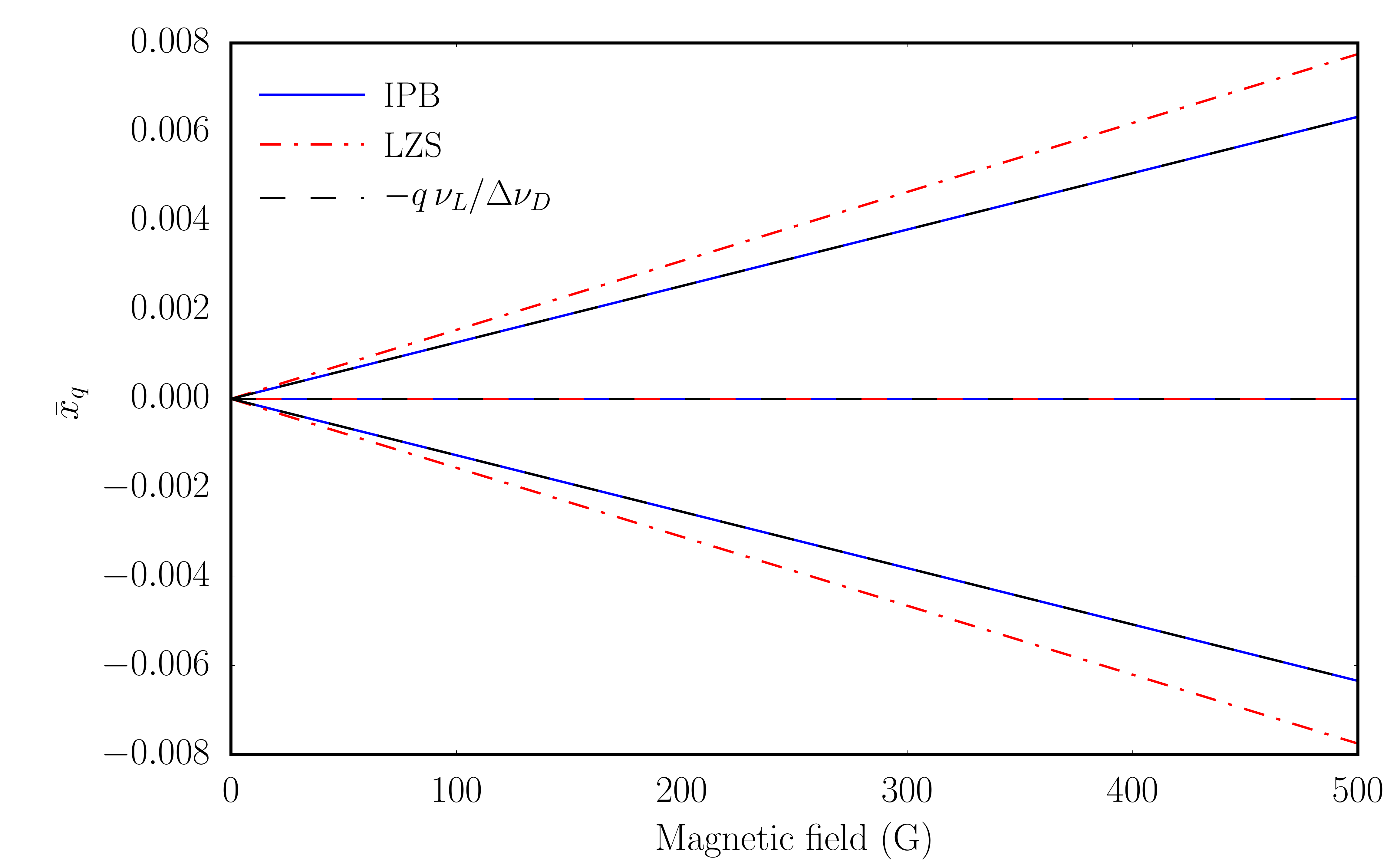}
 \caption{Spectral positions of the centers of gravity (see Equation~\eqref{FrequencyShifts}), taking a two-term atomic 
 model for the H~{\sc i} Ly$\alpha$ line. 
 The Doppler width has been taken at a height of $1998.5$~km in the FAL-C atmospheric model, corresponding to
  $54.4$~m\AA .  
The blue solid lines represent the spectral positions obtained in the incomplete Paschen-Back effect 
 regime, the red dashed-dotted lines represent the same values obtained under the linear Zeeman splitting approximation, 
 and the black dashed lines follow Equation~\eqref{FreqShiftNZT}. } 
 \label{FigAppendix2}
\end{figure} 
On the other hand, such shifts are considerably overestimated when the LZS approximation is made (see Figure~\ref{FigAppendix2}),
as a consequence of neglecting the magnetic dependence of the strengths of the various transitions. The necessity of fully 
accounting for the IPB effect in order to correctly
determine the spectral positions of the centers of gravity, also in the presence of magnetic fields weak enough that
the splitting they induce is much smaller than separation between FS levels, was already pointed out by 
 \cite{SocasNavarro+04}. 

The explicit expressions for the elements of the propagation matrix for a two-term atom with an unpolarized lower
term, in the presence of a magnetic field of arbitrary strength, can be
obtained as a particular case of those for a multi-term atom given in Section~7.6 of LL04. 
For the the purposes of this work, it is convenient to write such coefficients (defined taking the quantization axis parallel
to the magnetic field), for a given frequency $\nu$ and direction $\boldsymbol{\Omega}$ as
\begin{align}
 \eta_i(\nu,\boldsymbol{\Omega}) & = k_M \sum_{K} \sqrt{\frac{2 K + 1}{3}} \,  {\mathcal T}^K_0(i,\boldsymbol{\Omega}) \sum_q (-1)^{1+q}
 \left(\begin{array}{c c c}
        1  &  1  & K \\
        q  & -q & 0
       \end{array} \right)  \phi_q(\nu) \, , \quad \quad (i = 0, 1, 2, 3) \label{EtaMatQuant}\\
  \rho_i(\nu,\boldsymbol{\Omega}) & = k_M \sum_{K} \sqrt{\frac{2 K + 1}{3}} \, {\mathcal T}^K_0(i,\boldsymbol{\Omega}) \sum_q (-1)^{1+q}
 \left(\begin{array}{c c c}
        1  &  1  & K \\
        q  & -q & 0
       \end{array} \right)  \psi_q(\nu) \, ,  \quad \quad (i = 1, 2, 3)      
\label{RhoMatQuant}
       \end{align}
where $k_M$ is the so-called frequency-integrated absorption coefficient and ${\mathcal T}^K_0(i,\boldsymbol{\Omega})$
are the polarization tensors introduced in \cite{LandiDeglInnocenti83}. 
The $\phi_q$ and $\psi_q$ profiles appearing in the previous expression are given by 
\begin{equation}
\phi_q(\nu) \, = \sum_{j_u M_u j_\ell M_\ell} S^{j_u M_u, j_\ell M_\ell}_q \, \mbox{Re}\bigl\{\Phi(\nu_{j_u M_u, j_\ell M_\ell} - \nu) \bigr\} \, ; \quad \quad
\psi_q(\nu) \, = \sum_{j_u M_u j_\ell M_\ell} S^{j_u M_u, j_\ell M_\ell}_q \, \mbox{Im}\bigl\{\Phi(\nu_{j_u M_u, j_\ell M_\ell} - \nu) \bigr\} \, . 
  \label{PhiPsiProfs}
\end{equation}
In the proofs presented hereafter, we consider the observer's reference frame, making the assumption that the 
distribution of atomic velocities is Maxwellian. 
In terms of reduced frequencies, the complex absorption profiles $\Phi(\nu_{j_u M_u, j_\ell M_\ell} - \nu)$ introduced 
above can be given as 
\begin{equation}
 \Phi\bigl(\nu_{j_u M_u, j_\ell M_\ell} - \nu \bigr) = \frac{1}{\sqrt{\pi} \Delta\nu_D} \biggl(H\bigl(x + x_{j_u M_u, j_\ell M_\ell}, a\bigr)
 + \mathrm{i} \, L\bigl(x + x_{j_u M_u, j_\ell M_\ell},a\bigr) \biggr) \, , 
 \label{ProfRelIm}
\end{equation}
where $H$ and $L$ are the  
Voigt profile and the associated dispersion profile, respectively (see LL04 for their explicit expressions). They 
contain the damping parameter 
$a = \Gamma/(4 \pi \Delta\nu_D)$, 
where $\Gamma$ is the line-broadening parameter. 
Note that $\Gamma = \Gamma_R + \Gamma_E + \Gamma_I$, where $\Gamma_R$ is the radiative de-excitation rate, which corresponds
to the Einstein coefficient for spontaneous emission $A(\beta_u L_u S \rightarrow \beta_\ell L_\ell S)$, and $\Gamma_I$ and $\Gamma_E$ are 
the de-excitation rates due to inelastic and elastic collisions, respectively. 
The discussion presented below concerns frequencies far from line center, for which the condition 
$x^2 + a^2 \gg 1$ is fulfilled and so one can take the asymptotic expansion (see LL04) for the $H$ and $L$
 to the lowest order in $x$, 
\begin{equation}
H(x,a) \sim \frac{1}{\sqrt{\pi}} \frac{a}{x^2 + a^2} \, , \quad \quad \;   L(x,a) \sim \frac{1}{\sqrt{\pi}} \frac{x}{x^2 + a^2} \, .
  \label{AsympExpan}
\end{equation}
\section{B. Particular case: The two-level atom for a $0 - 1$ transition}
\label{App2L}
Before considering a two-term atomic model with arbitrary values for $L_u$, $L_\ell$, and $S$, let us first consider
the particular case in which $S = 0$ so that $J_u = L_u$ and $J_\ell = L_\ell$ (corresponding to the case of a
two-level atomic model). In a reference frame such that the quantization axis is taken along the direction of the magnetic field, the elements of the propagation matrix given in 
Eqs.~\eqref{EtaMatQuant} and \eqref{RhoMatQuant} can be written in a more compact form by introducing the generalized profile $\Phi^{K, K^\prime}_Q$ and the 
generalized dispersion profile $\Psi^{K, K^\prime}_Q$ 
\citep[e.g.,][]{LandiDeglInnocenti+91a}, yielding 
\begin{equation}
 \eta_i(x,\boldsymbol{\Omega}) = 
 k_M \sum_{K} {\mathcal T}^K_0(i,\boldsymbol{\Omega}) \, \Phi^{0,K}_0(J_\ell, J_u, x)  \, , \quad \quad 
\rho_i(x,\boldsymbol{\Omega}) = 
 k_M \sum_{K} {\mathcal T}^K_0(i,\boldsymbol{\Omega}) \Psi^{0,K}_0(J_\ell, J_u, x) \, . 
 \label{2lev}
 \end{equation}
By selecting the reference direction for positive Stokes $Q$ so that the $\eta_U$ and $\rho_U$ coefficients are zero, 
the previous expression can be given explicitly in terms of the angle $\alpha$ between the direction of propagation
and the magnetic field as
\begin{align}
& \eta_I(x,\boldsymbol{\Omega}) = k_M \, \Bigl(\Phi^{0,0}_0(J_\ell,J_u;x) + \frac{\sqrt{2}}{4} (3\cos^2\alpha - 1) \; \Phi^{0,2}_0(J_\ell,J_u;x)
  \Bigr) \, , \notag \\
& \eta_Q(x,\boldsymbol{\Omega}) = k_M \frac{3\sqrt{2}}{4} \sin^2\alpha \; \Phi^{\,0,\,2}_0(J_\ell,J_u;x) \, , 
\quad \quad \quad \rho_Q(x, \boldsymbol{\Omega}) = k_M \frac{3 \sqrt{2}}{4} \,\sin^2\alpha \; \Psi^{\,0,\,2}_0(J_\ell,J_u;x) \, . \notag \\ 
& \eta_V(x,\boldsymbol{\Omega}) = k_M \frac{\sqrt{6}}{2} \cos\alpha \; \Phi^{\,0,\,1}_0(J_\ell,J_u;x) \, , 
\; \quad \quad \quad \rho_V(x, \boldsymbol{\Omega}) = k_M \frac{\sqrt{6}}{2} \cos\alpha \; \Psi^{\,0,\,1}_0(J_\ell,J_u;x) \, .
\label{Prop2lev}
\end{align}
Taking also $J_u = 1$ and $J_\ell = 0$, as in the two-level model considered in previous sections, the 
generalized profiles and generalized dispersion profiles can be written as 
\begin{align}
 & \Phi^{\,0,\,0}_0(0,1;x) = \frac{1}{3} \, \Bigl[\phi_1 + \phi_0 + \phi_{-1} \Bigr] \, ,
 & \Phi^{\,0,\,1}_0(0,1;x) = \frac{\sqrt{6}}{6} \, \Bigl[\phi_1 - \phi_{-1} \Bigr] \, ,  \quad \quad
 & \Phi^{\,0,\,2}_0(0,1;x) = \frac{\sqrt{2}}{6} \, \Bigl[\phi_1 - 2 \phi_0 + \phi_{-1} \Bigr] \notag \\
 & \Psi^{\,0,\,0}_0(0,1;x) = \frac{1}{3} \, \Bigl[\psi_1 + \psi_0 + \psi_{-1} \Bigr] \, ,
 & \Psi^{\,0,\,1}_0(0,1;x) = \frac{\sqrt{6}}{6} \,  \Bigl[\psi_1 - \psi_{-1} \Bigr] \, , \quad \quad
 & \Psi^{\,0,\,2}_0(0,1;x) = \frac{\sqrt{2}}{6} \,  \Bigl[\psi_1 - 2 \psi_0 + \psi_{-1} \Bigr] \, .
\end{align}
We note that, for such a two-level atomic model, the transition strengths introduced  
in Equation~\eqref{TransStr} can simply be written as 
\begin{align}
 S^{M_u, M_\ell}_q = 3 \left(\begin{array}{c c c}
    J_u  & J_\ell  & 1 \\
    - M_u  & M_\ell & -q  
    \end{array} \right)^2 \, . 
 \label{2LevTRS}
\end{align}
Therefore, one can easily see that, in the case that $J_u = 1$ and $J_\ell = 0$, the profiles $\phi_q$ 
and $\psi_q$ given in Equation~\eqref{PhiPsiProfs} can also be given in the following, more compact, form: 
\begin{equation*}
 \phi_q = \frac{1}{\sqrt{\pi}\Delta\nu_D} H\bigl(x + \bar{x}_{q}, a \bigr) \, , \quad \quad
 \psi_q = \frac{1}{\sqrt{\pi}\Delta\nu_D} L\bigl(x + \bar{x}_{q}, a \bigr) \, .  
\end{equation*}
Using these expressions for the $\phi_q$ and $\psi_q$ profiles together with Equation~\eqref{AsympExpan}, valid when
considering spectral regions far from the line core, 
one can write the generalized profiles and generalized dispersion profiles as a sum of fractions of polynomials as
 \begin{align}
& \Phi^{0,0}_0(0,1;x) \sim \frac{a}{3 \pi \Delta\nu_D} \left[\frac{1}{a^2 + \bigl(x + \bar{x}_{1} \bigr)^2} + \frac{1}{a^2 + x^2} 
+ \frac{1}{a^2 + \bigl(x + \bar{x}_{-1} \bigr)^2} \right] \, , \\ 
& \Phi^{0,1}_0(0,1;x) \sim \frac{\sqrt{6} \,a}{6\pi\Delta\nu_D} \left[\frac{1}{a^2 + \bigl(x + \bar{x}_{1}\bigr)^2} - \frac{1}{a^2 + \bigl(x + \bar{x}_{-1}\bigr)^2} \right] \, , \\ 
& \Phi^{0,2}_0(0,1;x) \sim \frac{\sqrt{2} \,a}{6\pi\Delta\nu_D} \left[\frac{1}{a^2 + \bigl(x + \bar{x}_{1}\bigr)^2} - \frac{2}{a^2 + x^2} 
+ \frac{1}{a^2 + \bigl(x + \bar{x}_{-1}\bigr)^2} \right] \, , \\ 
& \Psi^{0,0}_0(0,1;x) \sim \frac{1}{3 \pi \Delta\nu_D} \left[\frac{x + \bar{x}_{1}}{a^2 + \bigl(x + \bar{x}_{1}\bigr)^2} + \frac{x}{a^2 + x^2} 
+ \frac{x + \bar{x}_{-1}}{a^2 + \bigl(x + \bar{x}_{-1}\bigr)^2} \right] \, , \\ 
& \Psi^{0,1}_0(0,1;x) \sim \frac{\sqrt{6}}{6\pi\Delta\nu_D} \left[\frac{x + \bar{x}_{1}}{a^2 + \bigl(x + \bar{x}_{1}\bigr)^2} - \frac{x + \bar{x}_{-1}}{a^2 + \bigl(x + \bar{x}_{-1}\bigr)^2} \right] \, , \\ 
& \Psi^{0,2}_0(0,1;x) \sim  \frac{\sqrt{2}}{6\pi\Delta\nu_D} \left[\frac{x + \bar{x}_{1}}{a^2 + \bigl(x + \bar{x}_{1}\bigr)^2} - \frac{2 x}{a^2 + x^2} 
+ \frac{x + \bar{x}_{-1}}{a^2 + \bigl(x + \bar{x}_{-1}\bigr)^2} \right] \, . 
\end{align}
Summing the various terms in the square parenthesis, each of the previous profiles can be expressed  
as a single ratio of polynomials. 
Taking the leading order in $x$ in the numerator and denominator, one reaches the following limits for their ratios over $\Phi^{0,0}_0(0,1;x)$, 
\begin{align*}
 \frac{\Phi^{0, 0}_0(0,1;x)}{\Phi^{0,0}_0(0,1;x)} = 1 \, ,
 & \quad \quad \quad \quad \quad \lim_{x \to \infty} \frac{\Phi^{0, 1}_0(0,1;x)}{\Phi^{0,0}_0(0,1;x)} \to 0 \, ,
 & \lim_{x \to \infty} \frac{\Phi^{0, 2}_0(0,1;x)}{\Phi^{0,0}_0(0,1;x)} \to 0 \, , \\
 \lim_{x \to \infty} \frac{\Psi^{0, 0}_0(0,1;x)}{\Phi^{0,0}_0(0,1;x)} \to \infty \, ,
 & \quad \quad \quad \quad \quad \lim_{x \to \infty} \frac{\Psi^{0, 1}_0(0,1;x)}{\Phi^{0,0}_0(0,1;x)} \to \frac{\sqrt{6}}{3} \frac{\nu_L}{a \Delta\nu_D} \, ,
 &  \lim_{x \to \infty} \frac{\Psi^{0, 2}_0(0,1;x)}{\Phi^{0,0}_0(0,1;x)} \to 0  \, .
\end{align*}
Thus, the only off-diagonal element of the propagation matrix {that}, divided by $\eta_I$, does not eventually fall to zero as $x$ increases is
$\rho_V/\eta_I$. This ratio instead reaches the constant value 
\begin{equation}
\lim_{x\to\infty} \frac{\rho_V(x,\boldsymbol{\Omega})}{\eta_I(x,\boldsymbol{\Omega})} \to \frac{4 \pi \nu_L \cos\alpha}{\Gamma} \, .
\label{FarWingRVEI}
\end{equation}
One immediate conclusion is that, far enough from the line center, the $\rho_V/\eta_I$ ratio is independent of the
Doppler width of the line and it scales linearly with the magnetic field strength. 
In the absence of collisions, 
$\Gamma$ simply becomes the 
Einstein coefficient for
spontaneous emission $A(\beta_u L_u S J_u \rightarrow \beta_\ell L_\ell S J_\ell)$. 
%Interestingly, the field strength for the onset of the Hanle effect is determined by the same parameters;
%the parameter that characterizes its efficacy for a two-level atom 
%is given by 
% $H_u = (2 \pi \nu_L g_u)/A(\beta_u L_u S J_u \rightarrow \beta_\ell L_\ell S J_\ell)$, for which the role played by collisions has also been neglected. 
Interestingly, {the} onset of the Hanle effect is {likewise} determined by {the ratio of the
Larmor frequency associated to the ambient magnetic field $\nu_L$ over the line-broadening parameter $\Gamma$}. {For a two-level atom the efficacy of the Hanle effect is characterized by the parameter} 
$H_u = (2 \pi \nu_L g_u)/A(\beta_u L_u S J_u \rightarrow \beta_\ell L_\ell S J_\ell)$, {where $g_u$ is the Land\'{e} factor of the upper level},  
{and for} which the role played by collisions has also been neglected. This illustrates why one should expect the modification of the scattering polarization signatures in the line core (due to the
Hanle effect) and in the line wings (produced by magneto-optical effects) to become significant at similar magnetic field 
strengths. Furthermore, the relation between the magnetic field strength (through the Larmor frequency) and $\bar{x}_q$, given in 
Equation~\eqref{FreqShiftNZT}, implies that $\nu_L = \bigl(\bar{\nu}_{-1} - \bar{\nu_1} \bigr)/2$. 
Thus, the far-wing limit given in Equation~\eqref{FarWingRVEI} can be directly related to the frequency separation between 
the centers of gravity of the $\sigma_b$ and $\sigma_r$ components as
\begin{equation}
  \lim_{x\to\infty} \frac{\rho_V(x,\boldsymbol{\Omega})}{\eta_I(x,\boldsymbol{\Omega})} \to 
 \frac{2 \pi \, \bigl(\bar{\nu}_{-1} - \bar{\nu}_{1} \bigr) \cos\alpha}{\Gamma} \, .
 \label{RhoLimIllust}
\end{equation}
\section{C. The two-term atom in the incomplete Paschen-Back regime}
\label{AppGeneral2T}
We can now generalize the results presented in Appendix.~\ref{App2L} to the case of a two-term atom with  
arbitrary values of $S$, $L_u$, and $L_\ell$, 
accounting for the IPB effect. 
Taking a reference frame for which the quantization axis is along the magnetic field direction, the elements of the propagation matrix
given in Eqs.~\eqref{EtaMatQuant} and \eqref{RhoMatQuant} can be rewritten 
as
\begin{flalign}
& \eta_I = k_M \biggl[\frac{\sqrt{3}}{3} \sum_q (-1)^{1+q} 
\left(\begin{array}{c c c}
       1  & 1  & 0\\
       q  & -q &  0
      \end{array}
 \right) \phi_q(x) 
 + \frac{\sqrt{30}}{12} \, \bigl(3 \cos^2\alpha - 1 \bigr)\,
 \sum_q (-1)^{1+q} 
 \left(\begin{array}{c c c}
        1  &  1  &  2 \\
        q  & -q  & 0
       \end{array} \right) \phi_q(x) \biggr] \, , \notag \\ %\label{etai_expl}
& \eta_V = k_M \frac{\sqrt{6}}{2} \, \cos\alpha \, \sum_q (-1)^{1+q}
 \left(\begin{array}{c c c}
        1  &  1  &  1 \\
        q  & -q  & 0
       \end{array} \right) \phi_q(x) \, , %\label{etav_expl} \\       
\quad \quad \;\, \eta_Q = k_M \frac{\sqrt{30}}{4}\, \sin^2\alpha \, \sum_q (-1)^{1+q}
 \left(\begin{array}{c c c}
        1  &  1  &  2 \\
        q  & -q  & 0
       \end{array} \right) \phi_q(x) \, , \notag \\
& \rho_V = k_M \frac{\sqrt{6}}{2} \, \cos\alpha \, \sum_q (-1)^{1+q}
 \left(\begin{array}{c c c}
        1  &  1  &  1 \\
        q  & -q  & 0
       \end{array} \right) \psi_q(x) \, , 
\quad \quad \; \rho_Q = k_M \frac{\sqrt{30}}{4}\, \sin^2\alpha \, \sum_q (-1)^{1+q}
 \left(\begin{array}{c c c}
        1  &  1  &  2 \\
        q  & -q  & 0
       \end{array} \right) \psi_q(x) \, . 
\end{flalign}
Considering a frequency far {enough} from the line center {that} the asymptotic expansion 
in Equation~\eqref{AsympExpan} can be applied to the absorption profiles, the $\phi_q$ and $\psi_q$ profiles become
\begin{equation*}
 \phi_q(x) = \sum_{r = 1}^N \frac{a}{\pi \Delta\nu_D} \, S^{r}_q \frac{1}{a^2 + \bigl(x + x_r \bigr)^2} \, , \quad \quad 
 \psi_q(x) = \sum_{r = 1}^N \frac{1}{\pi \Delta\nu_D} \, S^{r}_q \frac{x + x_r}{a^2 + \bigl(x + x_r \bigr)^2 } \, . \label{FarQB}
\end{equation*}
The label $r$ stands for the set of quantum numbers ($j_u$, $M_u$, $j_\ell$, $M_\ell$) {that correspond} to the transition between states $\ket{\beta_u L_u S j_u M_u}$ 
and $\ket{\beta_\ell L_\ell S j_\ell M_\ell}$ and $N$ is the total number of distinct transitions between the two terms. 
As in the derivation presented in the previous section, the ratios of polynomials appearing in the profiles can be summed into a single ratio. 
In order to obtain the expressions for the elements of the propagation matrix presented below, which are valid where $x \gg 1$, we have used the identities 
\begin{align*}
 \sum_{r=1}^N S^r_q = 1 \, , \;\quad \sum_{r=1}^N S^r_q \, x_r \equiv \bar{x}_q = -q \,\bigl(\nu_L/\Delta\nu_D\bigr) \, , \, \quad \sum_{s\ne r} x_s = - x_r \, , 
\end{align*}
We recall that the last equality in the second identity holds in the IPB effect regime, while the spectral shifts $\bar{x}_q$ 
are instead overestimated when the LZS approximation is made. 
We have also used the following useful relations for the Racah algebra $3j$ symbols: 
\begin{align*}
 & \sum_q (-1)^{1+q} 
 \left(\begin{array}{c c c}
  1  &  1   &  0 \\
  q  &  -q  &  0
 \end{array}\right) 
 = \sqrt{3} \, , 
 \quad \quad 
 \sum_q (-1)^{1+q} 
 \left(\begin{array}{c c c}
  1  &  1   &  1 \\
  q  &  -q  &  0
 \end{array}\right)
  = 0 \, , 
 \quad \quad \quad \;\;\;
 \sum_q (-1)^{1+q} 
 \left(\begin{array}{c c c}
  1  &  1   &  2 \\
  q  &  -q  &  0
 \end{array}\right) 
 = 0 \, , \\ 
 & \sum_q (-1)^{1+q} \, q 
 \left(\begin{array}{c c c}
  1  &  1   &  0 \\
  q  &  -q  &  0
 \end{array}\right) 
 = 0 \, , 
 \quad \quad
 \sum_q (-1)^{1+q} \, q 
 \left(\begin{array}{c c c}
  1  &  1   &  1 \\
  q  &  -q  &  0
 \end{array}\right)
  = \frac{\sqrt{6}}{3} \, , 
 \quad \quad
 \sum_q (-1)^{1+q} \, q 
 \left(\begin{array}{c c c}
  1  &  1   &  2 \\
  q  &  -q  &  0
 \end{array}\right) 
 = 0 \, . 
 \end{align*}
Taking only the leading orders in $x$ for both the numerator and denominator, after some tedious algebra one reaches the following expressions for the elements of the propagation matrix 
\begin{subequations}
 \begin{align}
&  \eta_I(x,\boldsymbol{\Omega}) \approx k_M \frac{1}{\pi \Delta\nu_D} \frac{a}{x^2} \, , \\
&  \eta_V(x,\boldsymbol{\Omega}) \approx k_M \frac{2\, a}{\pi \Delta\nu_D} \, \frac{1}{x^3} \, \frac{\nu_L}{\Delta\nu_D} \cos\alpha \, , 
\quad \quad \; \eta_Q(x,\boldsymbol{\Omega}) \approx k_M \frac{\sqrt{30}}{4 \pi \Delta\nu_D} \, \frac{a}{x^4} \sin^2\alpha \sum_q (-1)^{1+q} 
  \left(\begin{array}{c c c}
          1  &  1  & 2 \\
          q  & -q  & 0
       \end{array} \right) \mathrm{v}_q \, , \\
& \rho_V(x,\boldsymbol{\Omega}) \approx k_M \frac{1} {\pi \Delta\nu_D} \, \frac{1}{x^2} \, \frac{\nu_L}{\Delta\nu_D} \cos\alpha \, , 
\quad \quad \;  \rho_Q(\nu,\boldsymbol{\Omega}) \approx k_M \frac{\sqrt{30}}{4 \pi \Delta\nu_D} \, \frac{1}{x^3} \sin^2\alpha \sum_q (-1)^{1+q} 
  \left(\begin{array}{c c c}
         1  &  1  & 2 \\
         q  & -q  & 0
        \end{array} \right) \mathrm{w}_q \, ,  
 \end{align}
 \label{PropMatElemLeading}
\end{subequations}
where
\begin{equation}
 \mathrm{v}_q = \sum_{r=1}^N S^r_q \, \biggl(\sum_{s\ne r} x^2_s + 4 \sum_{s\ne r} \, x_s \sum_{\substack{t>s \\ t\ne r}} \, x_t \biggr) \, , \quad \quad \quad 
 \mathrm{w}_q = \sum_{r=1}^N S^r_q \, \biggl(\sum_{s\ne r} x^2_s + 4 \sum_{s\ne r} \, x_s \sum_{\substack{t>s \\ t\ne r}} \, x_t - 2 \, x^2_r \biggr) \, .
\end{equation}
It is immediate to realize that, also for a two-term atom with arbitrary values of $L_u$, $L_\ell$, and $S$, the only coefficient in the propagation matrix
whose ratio over $\eta_I$ does not fall to zero when $x \to \infty$ is $\rho_V$. Moreover, the expressions relating such ratio to the Larmor frequency and to 
the spectral distance between the centers of gravity of the $\sigma_b$ and $\sigma_r$ components are also recovered exactly as given in Eqs.~\eqref{FarWingRVEI} 
and \eqref{RhoLimIllust}, respectively. 
It {should be emphasized} that this proof is based on the relation $\bar{x}_q = -q \, \nu_L/\Delta \nu_D$, which is strictly 
valid in the IPB effect regime. In contrast, making the LZS approximation may introduce significant errors
in the determination of the far-wing value of the $\rho_V/\eta_I$ relation, even in the presence of relatively 
weak magnetic fields. 

\end{document}